\documentclass[a4paper]{jpconf}
\usepackage{bm}
\usepackage{graphicx}
\begin{document}
\title{Effect of the spin-orbit interaction and the electron phonon coupling on the electronic state in a silicon vacancy}

\author{Takemi Yamada, Youichi Yamakawa, Yoshiaki \=Ono}

\address{\it Department of Physics, Niigata University, Ikarashi, Japan}

\ead{takemi@phys.sc.niigata-u.acjp}

\begin{abstract}
The electronic state around a single vacancy in silicon crystal is
 investigated by using the Green's function approach. The triply
 degenerate charge states are found to be widely extended and account
 for extremely large elastic softening at low temperature as observed in
 recent ultrasonic experiments. When we include the LS coupling
 $\lambda_{\rm Si}$ on each Si atom, the 6-fold spin-orbital degeneracy
 for the $V^{+}$ state with the valence $+1$ and spin $1/2$ splits into
 $\Gamma_{7}$ doublet groundstates and $\Gamma_{8}$ quartet excited
 states with a reduced excited energy of $O(\lambda_{\rm Si}/10)$. We
 also consider the effect of couplings between electrons and Jahn-Teller
 phonons in the dangling bonds within the second order perturbation and
 find that the groundstate becomes $\Gamma_{8}$ quartet which is
 responsible for the magnetic-field suppression of the softening in
 B-doped silicon.
\end{abstract}

\section{Introduction}%%%%%%%%%%%%%%%%%%%%%%%%%%%%%%%%%%%%%%%%%%%%%%%%%%
%Intrinsic point defects in semiconductor, such as vacancy or
%interstitial, are influential factors to determine the quality of
%semiconductor devices. In particular, the direct observation of a single
%vacancy in the crystalline silicon (Si) has been one of the main issues
%in the region of semiconductor physics over a quarter of a century.

Recently, Goto {\it et al.} have succeeded in the direct observation of
the isolated vacancy in high-purity crystalline silicon (Si), with
extremely low vacancy concentration believed less than $10^{15}$
cm$^{-3}$, through the ultrasonic measurements at low
temperature\cite{ob-1,ob-2}. They have measured the elastic constants of
two type samples, non-doped Si, and boron (B)-doped Si and observed
reciprocal temperature dependence below 20 K down to 20 mK, called
'elastic softening', without the sign of the local
distortion\cite{ob-1}. The softening of non-doped Si is independent of
the external magnetic fields up to 10 T, while that of boron (B)-doped
Si is suppressed by the external magnetic fields of the order of 1
T. They have claimed that (1) the softenings are cause by the isolated
vacancy groundstates with 3-fold orbital degeneracy coupled to the
ultrasound and (2) non-doped Si has the non-magnetic neutral charge
state $V^0$ and B-doped Si has the charge state $V^+$ with the valence
$+1$ and the spin 1/2\cite{ob-1}. Based on the standard theory of
vacancy\cite{review1,review2,review3} established in 1980's, the
groundstate degeneracies for both $V^{0}$ and $V^{+}$ states are
resolved with symmetry lowering due to the Jahn-Teller distortion in
contrast to the newly observed softening. 

To understand such anomalous
softenings, new approaches focusing on the quantum many-body effect have
been performed on the basis of cluster model for the dangling-bond orbitals 
in a Si vacancy in a couple of years\cite{rs-1,rs-2,so}. 
Yamakawa {\it et al.}\cite{rs-1,rs-2} have revealed that the effect of
couplings between electrons and Jahn-Teller phonons as well as that of 
Coulomb interaction in the dangling bonds play crucial roles for the 
electronic states of both $V^{+}$ and $V^{0}$ states.
Matsuura and Miyake\cite{so} have studied the effects of spin-orbit 
interaction and the Coulomb interaction, and  have found that 
the $\Gamma_8$ quartet groundstate, 
which is responsible for the magnetic suppression of the softening 
in B-doped Si, is realized in the 
$V^{+}$ state by assuming a negative value of the spin-orbit coupling
on the vacancy state $\lambda_{\rm vc}<0$, although 
the explicit calculation for $\lambda_{\rm vc}$ has not been done so far. 

In the cluster model calculations mentioned above, the effect of the spatial extension of the vacancy state, which is important to determine the absolute value of the elastic softening, has not been considered. 
In our previous paper\cite{TY}, 
we have investigated the electronic state around a single
vacancy in infinite Si crystal on the basis of the Green's function
approach, and found that the $T_2$ triplet vacancy states 
are widely extended up to 20 {\AA} and are
responsible for the extreme enhancement of the Curie constant of the
quadrupole susceptibilities resulting in the elastic softening at low
temperature. The Curie constant for the trigonal mode 
is considerably larger than that for the
tetragonal mode as observed in the ultrasonic experiments\cite{ob-1}. 
The effect of the spin-orbit interaction, however, has not been considered there. 

The purpose of this paper is to elucidate the effect of the spin-orbit
interaction on the widely extended vacancy states. 
For this purpose, we extend our previous study\cite{TY} for the case with 
the LS coupling $\lambda_{\rm{Si}}$ on each Si atom, and then 
we calculate the spin-orbit splitting in the vacancy state $\Delta_{\rm vc}$ 
by using the Green's function approach. 
We also consider the effect of couplings between electrons and 
Jahn-Teller phonons in the dangling bonds within the second order
perturbation. 

%\newpage
\section{Model and Formulation}%%%%%%%%%%%%%%%%%%%%%%%%%%%%%%%%%%%%%%%%%
Our model Hamiltonian consists of the tight-binding (tb) Hamiltonian $H_{\rm{tb}}$
with LS-coupling on each Si site and a vacancy potential $H_{\rm{vc}}$,
and explicitly given by
\begin{eqnarray}
&&H=H_{\rm{tb}}+H_{\rm{vc}}, \\
&&H_{\rm{tb}}=\sum_{ij}\sum_{m m'}\sum_{\sigma} t_{ij}^{mm'} c^{\dagger}_{im\sigma}c_{jm'\sigma} 
+\sum_{i}\sum_{mm'}\sum_{\sigma\sigma'}\lambda_{\rm{Si}} M_{mm'}^{\sigma\sigma'} c^{\dagger}_{im\sigma}c_{im'\sigma'}
=\sum_{{\bf k}}\sum_{{\rm s}=1}^{16} \epsilon_{{\bf k} {\rm s}}c^{\dagger}_{{\bf k} {\rm s}}c_{{\bf k} {\rm s}}, \label{eq:Htb}\ \ \ \  \\ 
&&H_{\rm{vc}} = \Delta\sum_{m\sigma} c_{0m\sigma}^{\dagger}c_{0m\sigma},
\end{eqnarray}
where $c_{im\sigma}^{\dagger}$ is a creation operator for an
electron at site $i$ and orbital $m~(=3{\it s},3{\it p_x},3{\it p_y},3{\it p_z}$) with
spin $\sigma$, and $c_{{\bf k} {\rm s}}^{\dagger}$ is that for wave vector
${\bf k}$ and band ${\rm s}~(=1\sim 16$). 
In first term of eq. (\ref{eq:Htb}), the tb parameters $t^{mm'}_{ij}$ are written by the
Slater-Koster parameters and determined so as to fit the tb band
energies $\epsilon_{{\bf k}{\rm s}}$ to the LDA band energies\cite{LDA} for
the Si crystal, and explicitly given in Ref \cite{TY}. 
The second term in eq. (\ref{eq:Htb}) represents LS-coupling on each Si site 
with LS-coupling constant $\lambda_{\rm{Si}}$ which is experimentally
determined as $\lambda_{\rm Si}=29~{\rm meV}$\cite{LS}. 
Here, the spin-orbit matrix $\bm{M}$ is
\begin{eqnarray}
\bm{M}=\frac{1}{2}\bordermatrix{
& |p_x~\uparrow\rangle &|p_y~\uparrow\rangle &|p_z~\uparrow\rangle &|p_x~\downarrow\rangle &|p_y~\downarrow\rangle &|p_z~\downarrow\rangle \cr
\langle p_x~\uparrow    |& 0 & -i & 0 & 0 & 0 & 1 \cr
\langle p_y~\uparrow    |& i & 0 & 0 & 0 & 0 & -i \cr
\langle p_z~\uparrow    |& 0 & 0 & 0 & -1 & i & 0 \cr
\langle p_x~\downarrow  |& 0 & 0 & -1 & 0 & i & 0 \cr
\langle p_y~\downarrow  |& 0 & 0 & -i & -i & 0 & 0 \cr
\langle p_z~\downarrow  |& 1 & i & 0 & 0 & 0 & 0\cr
            }.
\end{eqnarray}
The vacancy potential $H_{\rm{vc}}$ excludes electrons from the vacancy
site by raising the energy levels $\Delta$ for the orbitals belong to the vacancy site. 
For $\Delta\longrightarrow\infty$, no electron exists at the vacancy site and then an effective vacancy state is realized.

%{Green's function:G}%%%%%%%%%%%%%%%%%%%%%%%%%%%%%%%%%%%%%%%%%%%%%%%%%%%
In the absence of the vacancy ($\Delta=0$), the Green's function for the
perfect crystal is described as
\begin{equation}
G^{0\alpha\beta}_{ij}(z)=\sum_{{\bf k}{\rm s}}\frac{u_{\alpha {\rm s}}({\bf k})u_{\beta {\rm s}}^{*}({\bf k})}{z-\epsilon_{{\bf k}{\rm s}}}e^{i{\bf k}\cdot\left({\bf r}_i-{\bf r}_j\right)}\label{eq:G0}
\end{equation}
where $u_{\alpha {\rm s}}({\bf k})$ is the eigenvector for the energy
band $\epsilon_{{\bf k}{\rm s}}$ given in eq. (\ref{eq:Htb}) and orbital $m$ with spin $\sigma$, $\alpha=(m,\sigma)$. 
In the presence of the vacancy ($\Delta\ne 0$), the Green's function 
is obtained by solving the Dyson's equations which can be written in the $8\times8$ matrix representation as
${\rm{\bf G}}_{ij}={\rm{\bf G}}^{0}_{ij}+{\rm{\bf G}}^{0}_{i0}{\bf\Delta}{\rm{\bf G}}_{0j} $,
with the vacancy potential matrix 
$({\bf\Delta})_{\alpha\beta}=\Delta\delta_{\alpha\beta}$, 
where $({\rm{\bf G}}^0_{ij})_{\alpha\beta}=G^{0\alpha\beta}_{ij}$ 
is the Green's function for $\Delta=0$ given in eq. (\ref{eq:G0}) 
and $({\rm{\bf G}}_{ij})_{\alpha\beta}=G^{\alpha\beta}_{ij}$ is 
the corresponding Green's function for $\Delta\ne 0$. 
In the limit $\Delta\to \infty$, ${\rm{\bf G}}_{ij}\to 0$ with $i=0$ 
and/or $j=0$, and then
${\rm{\bf G}}^{0}_{0j}+{\rm{\bf G}}^{0}_{00}{\bf \Delta}{\rm{\bf G}}_{0j} \rightarrow 0$.
By using this relation, ${\rm{\bf G}}_{ij}$ is obtained by ${\rm{\bf G}}^{0}$ as
${\rm{\bf G}}_{ij}={\rm{\bf G}}^{0}_{ij}-{\rm{\bf G}}^{0}_{i0}\left({\rm{\bf G}}^{0}_{00}\right)^{-1}{\rm{\bf G}}^{0}_{0j}$.

%\newpage
\section{Results}%%%%%%%%%%%%%%%%%%%%%%%%%%%%%%%%%%%%%%%%%%%%%%%%%%%%%%
We calculate $G^{0\alpha\beta}_{ij}$ in eq. (\ref{eq:G0}) by
performing the $\rm{\bf k}$ summation with up to $20\times 20\times 10 =4000$
mesh points and obtain ${\rm{\bf G}}_{ij}$.  When
$\lambda_{\rm Si}=0$, a remarkable localized level is found in
the band gap and the total weight of this state is found to be 6 by
summing the contribution to this level from all sites. Therefore, we
find that this level corresponds to the $T_2$ states with the 6-fold
spin-orbital degeneracy with the energy $E_{T_2}$. This localized level is occupied by two
electrons in the $V^0$ state and by one electron in the $V^+$ state, respectively. 

\begin{figure}[t]
\begin{minipage}{0.45\hsize}
\vspace{-20mm}
\begin{center}
\includegraphics[width=7.5cm]{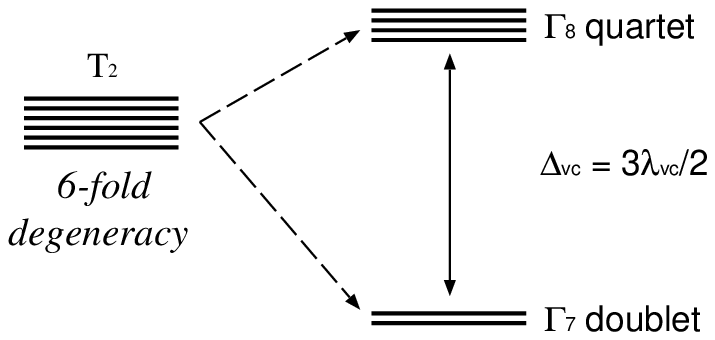}
\caption{Spin-orbit splitting in the vacancy state $\Delta_{\rm vc}$ for the case with a positive value of the spin-orbit coupling on the vacancy state $\lambda_{\rm vc}>0$.}
\label{fig:so1}
\end{center}
\end{minipage}
\hspace{0.5cm}
\begin{minipage}{0.5\hsize}
\begin{center}
\includegraphics[width=7cm]{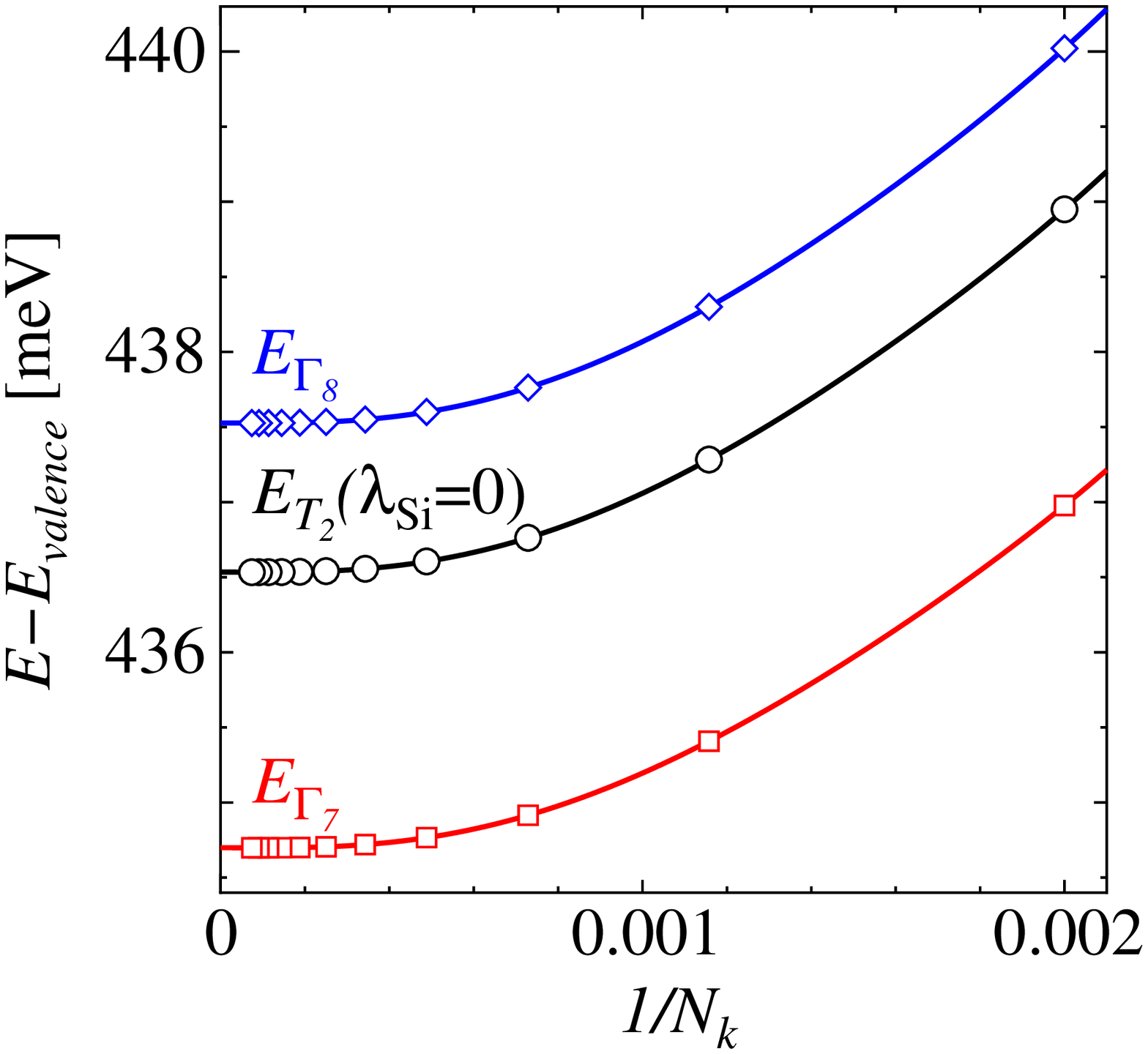}
\caption{Vacancy levels $E_{\Gamma_8}$  (\opendiamond) and $E_{\Gamma_7}$ 
(\opensquare) for $\lambda_{\rm Si}=43$ meV together with 
$E_{T_2}$ (\opencircle) for $\lambda_{\rm Si}=0$ 
as functions of the
 inverse of total number of $\rm{\bf k}$-points $1/N_{k}$.}
\label{fig:so2}
\end{center}
\end{minipage}
\end{figure}

By introducing the LS coupling on each Si atom ($\lambda_{\rm{Si}}=29~{\rm meV}$), the $T_2$ states split into
$\Gamma_{7}$ doublet groundstates and $\Gamma_{8}$ quartet excited
states with a spin-orbit splitting on the vacancy states $\Delta_{\rm
vc}>0$ as shown in Fig.\ref{fig:so1}. In Fig.\ref{fig:so2}, 
the energy levels for the vacancy states, $E_{T_2}$, $E_{\Gamma_8}$ and 
$E_{\Gamma_7}$ measured relative to the valence band top, are plotted as
functions of the inverse of total number of k-points $1/N_{k}$. Then, we
obtain a $N_k\rightarrow\infty$ extrapolated value of the spin-orbit
splitting on the vacancy states as $\Delta_{\rm vc}=2.83~{\rm meV}$
which is of order of $1/10$ of the spin-orbit splitting on Si atom
$\Delta_{\rm Si}= 43~{\rm meV}$\cite{LS}. Therefore, the spin-orbit
coupling on the vacancy states $\lambda_{\rm vc}$ becomes extremely
small as compared to $\lambda_{\rm Si}$ but is positive only due to the
effect of spatial extension of the vacancy state.

%%%%%%%%%%%%%%%%%%%%%%%%%%%%%%%%%%%%%%%%%%%%%%%%%%%%%%%%%%%%%%%%%%%%%%%%
%%%%%%%%%%%%%%%%%%%%%%%%%%%%%%%%%%%%%%%%%%%%%%%%%%%%%%%%%%%%%%%%%%%%%%%%

%\newpage 
Finally, we consider the effect of couplings between electrons and
nonadiabatic Jahn-Teller phonons in the dangling bonds by using the
Hamiltonian previously studied in Ref \cite{rs-2}, which is an extention 
of the early 
Schl${\rm{\ddot u}}$ter's model \cite{review3}. Within the second order
perturbation theory, we find that the groundstate becomes $\Gamma_{8}$
quartet for a realistic value of the electron-phonon coupling constant.
The explicit results will be shown in a subsequent paper\cite{TY2}. 
%This means that taking into account the electron phonon coupling is
%significant for explaining the observed suppression of the softening in B-doped Si.
%Therefore, the observed suppression of the softening in B-doped Si is
%accountable by taking into account the electron phonon coupling together
%with the both effect of the spin-orbit coupling on the vacancy states
%and the spatial extension.

\section{Summary and Discussion}%%%%%%%%%%%%%%%%%%%%%%%%%%%%%%%%%%%%%%%%
In summary, we have investigated the electronic state around a single
vacancy in infinite Si crystal on the basis of the Green's function
approach. When the LS coupling $\lambda_{\rm Si}$ on each Si atom is
taken into account, the $T_2$ states with the 6-fold spin-orbital degeneracy for
the $V^{+}$ state split into $\Gamma_{7}$ doublet groundstates and
$\Gamma_{8}$ quartet excited states with a reduced splitting energy of
$O(\Delta_{\rm Si}/10)$. We have also considered the effect of couplings
between electrons and Jahn-Teller phonons in the dangling bonds within
the second order perturbation and find that the groundstate becomes
$\Gamma_{8}$ quartet which is responsible for the magnetic-field
suppression of the softening in B-doped Si.

In the present study, we have assumed that the $T_d$-point symmetry is preserved in a single vacancy in infinite Si crystal. 
Even in this case, the lattice relaxation with keeping the $T_d$-point symmetry  might take place. 
Therefore, we have also investigated the effect of symmetry preserving lattice relaxation on the spin-orbit splitting in the vacancy states $\Delta_{\rm vc}$ by using the first-principle calculation with supercell method, and have found that $\Delta_{\rm vc}$ is positive in a single vacancy with an unitcell with 63 Si atoms \cite{TY2}. To be more conclusive, we need further investigation to determine the explicit value of $\Delta_{\rm vc}$ with a larger size of supercell.

The effect of the electron correlation, which is considered to be important 
to determine the many-body groundstate in a Si vacancy as discussed from the cluster model calculations \cite{rs-1,rs-2,so}, has not been considered in this paper.  The effect can be discussed by including the selfenergy corrections on the basis of the present Green's function approach and will be discussed in near future. 

\section*{Acknowledgments}
The authors thank T. Goto, H. Kaneta, Y. Nemoto, K. Mitsumoto, K. Miyake
and H. Matsuura for many useful comments and discussions. 
This work was partially supported by the Grant-in-Aid for Scientific
Research from the Ministry of Education, Culture, Sports, Science and Technology.
\\

\end{document}